\documentclass[runningheads,a4paper]{llncs}
\pdfoutput=1

\usepackage{url}
\usepackage{breakcites}
\usepackage{ textcomp }
\usepackage{ comment }
\usepackage[color=yellow!60,textsize=footnotesize,obeyDraft,draft]{todonotes}
\usepackage{colortbl}
\usepackage{booktabs}
\usepackage{capt-of}
\usepackage{makecell}
\usepackage{textcomp }
\usepackage{multirow}
\usepackage{listings}
\usepackage{graphicx}
\usepackage{colortbl}
\usepackage{booktabs}
\usepackage{amsmath}
\usepackage[font=footnotesize,labelfont=bf]{subcaption}
\usepackage[hidelinks]{hyperref}
\usepackage[capitalize]{cleveref}
\usepackage{tikz}
\usepackage{pgfplots}
\usepackage[para,online,flushleft]{threeparttable}
\usepackage{multirow}
\usepackage{wasysym}
\usepackage{amssymb}
\usetikzlibrary{arrows}

\newcounter{mnotei}
\newcolumntype{L}[1]{>{\raggedright\let\newline\\\arraybackslash\hspace{0pt}}m{#1}}
\newcolumntype{C}[1]{>{\centering\let\newline\\\arraybackslash\hspace{0pt}}m{#1}}
\newcolumntype{R}[1]{>{\raggedleft\let\newline\\\arraybackslash\hspace{0pt}}m{#1}}

\newcommand{\includegraphicsmaybe}[2]{
    \IfFileExists{#2}{\includegraphics[#1]{#2}}{
    \detokenize{File #2 is missing, maybe you need to run plots.py?}
}}

\begin{document}
\bibliographystyle{ieeetr}
\mainmatter

\title{On the Value and Limits of Multi-level Energy Consumption Static Analysis for Deeply Embedded Single and Multi-threaded Programs}

\titlerunning{Energy Consumption Static Analysis}

\author{Kyriakos Georgiou, Steve Kerrison, Kerstin Eder}

\authorrunning{K. Georgiou et al.}
\institute{University of Bristol}
\tocauthor{Authors' Instructions}
\maketitle

\begin{abstract}
There is growing interest in lowering the energy consumption of computation. Energy transparency is a concept that makes a program's energy consumption visible from software to hardware through the different system layers. 
Such transparency can enable energy optimizations at each layer and between layers, and help both programmers and operating systems make energy aware decisions. 
The common methodology of extracting the energy consumption of a program is through direct measurement of the target hardware. This usually involves specialized equipment and knowledge most
programmers do not have. 
In this paper, we examine how existing methods for static resource analysis and energy modeling can be utilized to perform Energy Consumption Static Analysis (ECSA) for deeply embedded programs. 
%
%
To investigate this, we have developed ECSA techniques that work at the instruction set level and at a higher level, the LLVM IR, through a novel mapping technique.
We apply our ECSA to a comprehensive set of mainly industrial benchmarks, including single-threaded and also multi-threaded embedded programs from two commonly used concurrency patterns, task farms and pipelines. 
We compare our ECSA results to hardware measurements and predictions obtained based on simulation traces.
We discuss a number of application scenarios for which ECSA results can provide energy transparency and conclude with a set of new research questions  for future work.
\end{abstract}

\section{Introduction}
\label{sec:introduction}

A substantial amount of effort has been invested into predicting the execution time of a program.  However, there is little in the complementary area of energy consumption. Such information can be of significant value during the development and life time of critical systems. For example, energy consumption information can be crucial for devices that depend on unreliable, limited sources of power such as energy harvesters. Giving consideration to the energy consumption of a system at development time can avoid potential system failures due to inadequate energy supply at runtime. For systems that operate on a battery, this can provide a good approximation of the time frame in which the battery needs replacement.

The energy consumption of a program on specific hardware can always be determined through physical measurements. Although this is potentially the most accurate method, it is often not easily accessible. Measuring energy consumption can involve sophisticated equipment and special hardware knowledge. Custom modifications may be needed to probe the power supply. These conditions
make it very difficult for the majority of software developers to assess a program's energy consumption.

Static Resource Analysis (SRA) provides an alternative to measurement. Significant progress has been made in the area of Worst Case Execution Time (WCET) prediction using static techniques that determine safe upper bounds for the execution time of programs. This naturally leads to the question of whether similar techniques can be used to bound the energy consumption of programs, and, if so, how effective they can be. A popular approach used for WCET is the Implicit Path Enumeration Technique (IPET), which retrieves the worst case control flow path of programs based on a timing cost model. Instead, in~\cite{Jayaseelan2006}, an energy model that assigns energy values to blocks of Instruction Set Architecture (ISA) code is used, and the authors claim to statically estimate Worst Case Energy Consumption (WCEC).

However, in contrast to timing, energy consumption is data sensitive, i.e.\ the energy cost of executing an instruction varies depending on (the circuit switching activity caused by) the operands used. This effect is not captured in non data sensitive energy models, i.e.\ models that assign a single energy consumption value to each entity, e.g.\ to each instruction. Such models typically are characterized based on averages obtained from measuring the energy consumed when random data is being processed~\cite{Tiwari1996}. Alternatively, the highest energy consumption measured could be used for model characterization. As a consequence, when a non data sensitive energy model is used, the safety of the bounds retrieved from a worst case path static analysis might be undermined by worst case data scenarios for models that provide average energy consumption costs. On the other hand, the use of worst case models is known to lead to over-estimations~\cite{DBLP:journals/corr/PallisterKME15} affecting the tightness of the retrieved bounds, because it is unlikely that the data that triggers the worst case energy consumption for one instruction also does this for all subsequent instructions in a program.
This problem applies to all previous works that perform static analysis for
energy consumption, as they combine non data sensitive bound analysis
techniques with non data sensitive energy models.
In \cite{Jayaseelan2006} static analysis for WCEC is claimed by maximising the switching activity factor for each simulated component. However, the model abstraction level used does not guarantee that a physical implementation would behave in this way. We use a model with a similar constraint, where the data input that would trigger the worst case per instruction is not known, and so cannot assert the results to be WCEC.

In this paper we thoroughly investigate the value and limitations of using IPET
in combination with non data sensitive energy models to perform Energy
Consumption Static Analysis (ECSA) in the context of deeply embedded hardware,
in our case the XMOS XS1-L ``Xcore''~\cite{XMOS2009a}. 
The Xcore is a multi-threaded deeply embedded processor with time-deterministic instruction execution. Such systems are simpler than general purpose processors and favor predictability and low energy consumption over maximizing performance. The absence of performance enhancing complexity at the hardware level, such as caches, provides us with an ideal setting to evaluate ECSA.

We base our investigation on an ISA-level multi-threaded energy model for the Xcore~\cite{Kerrison15}. This model was characterized using constrained pseudorandom input data and associates a single averaged energy cost with each instruction in the XMOS ISA. 
We refined this model to one that is well suited for ECSA as it represents both static and dynamic power contributions to better reflect inter-instruction and inter-thread overheads; this improved model accuracy by an average of 4\%. 
In addition to using this model for ECSA, we also used it to compare ECSA results with predictions based on statistics obtained from simulation traces.

For our study we have developed an IPET-based ECSA, which we use together with the non data sensitive ISA-level energy model described above, to predict the energy consumption of single and multi-threaded programs. With respect to the latter we focus on two commonly used concurrency patterns in embedded programs, task farms and pipelined programs with evenly distributed workloads across threads. 
In addition, we have developed a novel mapping technique to lift our ISA-level energy model to a higher level, the intermediate representation of the compiler, namely LLVM IR~\cite{LattnerLLVM2004}, implemented within the LLVM tool chain~\cite{LLVM}.
This enables ECSA to be performed at a higher level than ISA, thus introducing
energy transparency into the compiler tool chain by making energy consumption
information accessible directly to the optimizer.

Performing ECSA on multi-threaded programs and at the LLVM IR allows a comprehensive analysis of the energy consumption predictions that can be obtained using this technique. 
Our ECSA technique is evaluated using a set of single- and multi-threaded
benchmarks, mainly selected from a number of industrial embedded applications.
Our results show that accurate energy estimations can be retrieved at the ISA level. The mapping technique allowed for energy consumption transparency at the LLVM IR level,
with accuracy keeping within 1\% of ISA-level estimations in most cases.
\noindent The main contributions of this paper are:
\begin{enumerate}
\item Modeling the target architecture to capture its behavior statically,
  including refinement of an existing ISA-level energy model, improving its
  accuracy by around 4\% (\Cref{sec:xs1model});
\item Formalization and implementation of a novel mapping technique that lifts
  an ISA-level energy model to a higher level, the intermediate representation of the LLVM
  compiler, which allows ECSA of programs at the LLVM IR level
  (\Cref{subsec:mapping});
\item ECSA on a set of multi-threaded programs
  (\Cref{subsec:mult-threadedAnalysis}), focusing on task farms and pipelines,
  two commonly used concurrency patterns in embedded computing;
\item Comprehensive evaluation of our ECSA on a set of industrial benchmarks
  and detailed analysis of results (\Cref{sec:results});
\item Discussion of the practical value and limitations of how such analysis
  can be useful for software developers, compiler engineers, development tools
  and Real Time Operating Systems (RTOS) (\Cref{ECSAapplications}).
\end{enumerate}

The rest of the paper is organized as follows. \Cref{sec:background} critically reviews previous work on energy modeling and SRA, with a focus on SRA for energy consumption and the effects of combining non data sensitive bound analysis techniques with non data sensitive energy models. \Cref{sec:SAnalysis} introduces in detail the components of our analysis, in particular the formalization and implementation of our mapping technique, and how ECSA can be applied to multi-threaded programs. Our experimental evaluation methodology, benchmarks and results are presented and discussed in \Cref{sec:evaluation}. \Cref{sec:conc_future} concludes the paper, outlines opportunities for future work and raises a number of research questions to stimulate further research in ECSA.

\section{Background}
\label{sec:background}

The work presented in this paper builds upon two areas: processor energy modeling and SRA. This section establishes the background work of both.

\subsection{Energy modeling of embedded processors}
\label{sec:modeling_review}

Energy modeling can be performed at various levels of abstraction, from gate-
or transistor-level in detailed hardware simulation~\cite{Bogliolo1997}, up to
high-level modeling of whole applications. Although the hardware components are
responsible for power dissipation and thus consumption of energy, the behavior
of that hardware is largely controlled by the software running upon it. As
such, writing software that makes efficient use of the underlying hardware has
been identified as the most important step in energy efficient software
development~\cite{Roy1997}. For energy modeling to be useful to a software
developer, models must convey information that can be related to the code
the developer is writing.

The ISA is a practical level of abstraction for energy modeling of software,
because it expresses the underlying operations performed by the hardware and
its relationship with the intent of the software. In~\cite{Tiwari1996} an
ISA-level energy model is proposed that obtained energy consumption data
through hardware measurements of large loops of individual instructions. The
total cost of a program is composed of instruction costs, inter-instruction
costs (the effects of switching from one instruction to the next), and
externally modeled behaviour such as activity in the memory hierarchy.

This work was initially applied to x86 and SPARC architecture processors, operating with an accuracy of within 10\% of the hardware. It was extended to form a framework  for architecture-level power analysis, Wattch~\cite{Brooks2000}. The Sim-Panalyser~\cite{Simpanalyzer} uses a similar approach, built on top of the SimpleScalar architecture simulation framework~\cite{Austin2002}.

If additional characteristics of processor activity are considered, such as
bit-flips in the data-path, a more accurate data-dependent model can be
produced, such as that of~\cite{Steinke2001,Sarta1999}. This requires more
detailed information from simulation in order to supply additional model
parameters, but has been demonstrated to bring accuracy to within 1.7\% of the
hardware. It is still an abstraction away from the internal switching activity
of functional units, however. Observing the results in~\cite{Kojima1995}, some
functional units may be more dependent on their internal structure than
input/output Hamming weight with respect to data-dependent power.

Using similar approaches to Steinke and Tiwari, additional processor architectures such as VLIW DSPs have also been modeled, with 4.8\%~\cite{Sami2002} and 1.05\%~\cite{Ibrahim2008} accuracy.  Alternative approaches to modeling include representing activity in terms of the processor's functional blocks~\cite{Ibrahim2008}, energy profiling of the most commonly used software library functions~\cite{Qu2000}, and construction of model parameters through linear regression~\cite{LeeRegression2001}.

In~\cite{Jayaseelan2006}, a micro-architectural energy model was created, considering functional units activity, clock gating and pipeline progression for a simulated processor. This model was used for WCEC static analysis. To retrieve safe bounds, the switching activity factor was set to the maximum, 1.0, for each component. This led to significant energy consumption over-estimations in some cases, up to 33\%, and assumes that the model accurately reflects a physical implementation.

In architectures where performance counters are available, these can be used to characterize the processor energy consumption based on the conditions affecting these counters, such as cache misses and pre-fetches. Simulations that model these performance events can then be used to predict the energy consumption of an application. This has been applied to processors of various
levels, from embedded XScale~\cite{xscaleunitevents} to Xeon Phi accelerators~\cite{phimodel}.

The discussed approaches achieve varying levels of accuracy, all within a 10\% error margin. The comparison points vary between methods, so the accuracies are not necessarily directly comparable. However, the prior work motivates new models to achieve a similar margin. In many of the above examples, the models target a `typical' energy characterization, where the modeled energy
consumption is based on random or non-exhaustive input data sets. For a given application, some additional error margin will be introduced based on the particular characteristics of its dataset. This forms a part of the model error, in addition to the errors arising from the abstractions applied in each model type. The work presented in this paper, which examines multiple
abstraction levels, seeks to identify each point at which inaccuracies may be introduced into the estimation process. This is important to assesses usefulness of estimations produced by static analysis, and will be discussed in \Cref{sec:results}

\subsection{Static Resource Analysis}
\label{subsec:SRAbackground}

SRA is a methodology to determine the usage of a resource (usually time or energy or both) for a specific task when executed on a piece of hardware, without actually executing the task. This requires accurate modeling of the hardware's behavior in order to capture the dynamic
functional and non-functional properties of task execution. Determining these properties accurately is known to be undecidable in general. Therefore, to extract safe values for the resource usage of a task, a sound approximation is needed~\cite{Wilhelm:2008,brat2014ikos}.

SRA has been mainly driven by the timing analysis community. Static cost analysis techniques based on setting up and solving recurrence equations date back to Wegbreit's~\cite{Wegbreit:1975} seminal paper, and have been developed significantly in subsequent work~\cite{Rosendahl:1989,Debray:1990,Debray:1997,Vasconcelos:2003,Navas:2007,Albert2011}. Other classes of approaches to cost analysis use dependent types~\cite{DBLP:journals/toplas/0002AH12}, SMT solvers~\cite{Alonso2012}, or size change abstraction~\cite{DBLP:journals/corr/abs-1203-5303}.

For performing an accurate WCET static analysis, there are four essential components~\cite{Wilhelm:2008}:

\begin{enumerate}
\item Value analysis: mainly used to analyze the behavior of the data cache.
\item Control flow analysis: used to identify the dynamic behavior of a program.
\item Low level or processor behavior analysis: attempts to retrieve timing costs for each atomic unit on a given hardware platform, such as an instruction or a basic block (BB) in a Control Flow Graph (CFG) for a processor.
\item Calculation: uses the results from the two previous components to estimate the WCET. Most common techniques used for calculation of the WCET are the IPET, the path-based techniques and the tree-based methods~\cite{Engblom:2000a}.
\end{enumerate}

Three of the above components, namely the control flow analysis, low level analysis and calculation, are adopted in our work and will be further explained in \Cref{sec:SAnalysis}.

IPET is one of the most popular methods used for WCET analysis \cite{Li:1995b,Theiling:1998,Ottosson:1997,Engblom:2000a,PotopButucaru:2013}. In this approach, the CFG of a program is expressed as an Integer Linear Programming (ILP) system, where the objective function represents the execution time of the program. The problem then becomes a search for the WCET by maximizing the retrieved objective function under some constraints on the execution counts of the CFG's basic blocks. The main advantage of this technique is the ability to determine the basic blocks in the worst case execution path and their respective execution counts without the need to extract the explicit worst execution path (ordered list of the executed basic blocks). This is more efficient than path based techniques for retrieving WCET bounds~\cite{Engblom:2000a}.

In the presence of caches or a complex processor pipeline, the ILP solving complexity can increase dramatically, making IPET not practical for WCET. Abstract interpretation~\cite{CousotCousot:1977}, a technique used to facilitate data flow analysis, can then be used in conjunction with IPET to allow WCET in such cases~\cite{Theiling:1998}.

Although significant research has been conducted in static analysis for the execution time estimation of a program, there is little on energy consumption. One of the few approaches~\cite{Navas:2008} seeks to statically infer the energy consumption of Java programs as functions of input data sizes, by specializing a generic resource analyzer~\cite{Navas:2007,Hermenegildo:2005} to Java bytecode analysis~\cite{Navas:2009}. However, a comparison of the results to actual
measurements was not performed.
Later, in~\cite{Liqat:2014}, the same generic resource analyzer was instantiated to perform energy analysis of XC~programs~\cite{Watt:2009} at the ISA level based on ISA-level energy models and including a comparison to actual hardware measurements. However, the scope of this particular analysis approach was limited to a small set of simple benchmarks because information required for the analysis of more complex programs, such as program structure and types, is not available at the ISA level. The analysis presented in this paper does not rely on such information. 
A similar approach, using cost functions, was used in~\cite{grech15}. The analysis was performed at the LLVM IR level, using the mapping technique that we formalise and describe in full detail for the first time in this paper. Although the range of programs that could be analyzed was improved compared to~\cite{Liqat:2014}, the complexity of solving recurrence equations for analysing larger programs proved a limiting factor.

In~\cite{Jayaseelan2006} the WCEC for a program was inferred by using the IPET first introduced in~\cite{Li:1995b}. They claim WCEC analysis, and experimental results indicate that all energy estimations over-approximate the energy consumptions retrieved from simulation. However, infeasible paths were not excluded from analysis, and there is no guarantee that the comparison test cases used in simulation were the actual worst cases.

Similarly, in~\cite{wagemannworst} the authors attempt to perform static worst case energy consumption analysis for a simple embedded processor, the \texttt{Cortex M0+}. This analysis is also based on IPET combined with a so called absolute energy model, an energy model that is said to provide the ``maximum energy consumption of each instruction''. The authors argue that they can retrieve a safe bound. However, this is demonstrated on a single benchmark, {\tt bubblesort}, only. The bound is 19\% above a single hardware measurement; the authors acknowledge that this approach leads to over-approximations. Furthermore, the hardware measurement used as a base line to evaluate the prediction obtained from static analysis only captures the algorithm's worst case complexity scenario, no information is given on the actual data to provide insight into the effect of data switching activity on energy consumption. This can be misleading, since two sets of different input data might have the same algorithmic worst case behavior, but can be very different with respect to their total energy consumption. In practice, this gives rise to a range of energy consumption measurements for different input data all triggering the algorithmic worst case path. For instance, for the Xcore architecture the energy consumption of the \texttt{MatMult 4} threads benchmark~\cite{benchmarks} for the same size of matrices, ranges from 4.1 to 4.9\,nJ depending on the used data. We have closely investigated this and discuss our findings in \Cref{dataEffect}.

All of the reviewed previous works for static energy consumption analysis used worst case path analysis methods combined with non input pattern-depended energy models. Currently, there is no practical method to perform average case static analysis~\cite{townley2013practical}. One of the most recent works towards average case SRA, demonstrates that compositionality combined with the capacity for tracking data distributions unlocks the average case analysis, but novel language features and hardware designs are required to support these properties~\cite{Schellekens201061}.
Furthermore, developing a data sensitive energy model requires detailed knowledge of and access to the RTL, since the power dissipation is highly depended on the switching activity inside the circuits~\cite{Najm1994}; this is a challenge in itself. This situation has motivated us to conduct a comprehensive study to fully understand the value and limitations of ECSA, using IPET-based analysis in combination with a single cost energy model, for both single and multi-threaded code at the ISA and LLVM IR levels of abstraction.

\section{Energy Consumption Static Analysis}
\label{sec:SAnalysis}

\begin{figure*}[ht]
\centering
\includegraphics[width=0.75\textwidth]{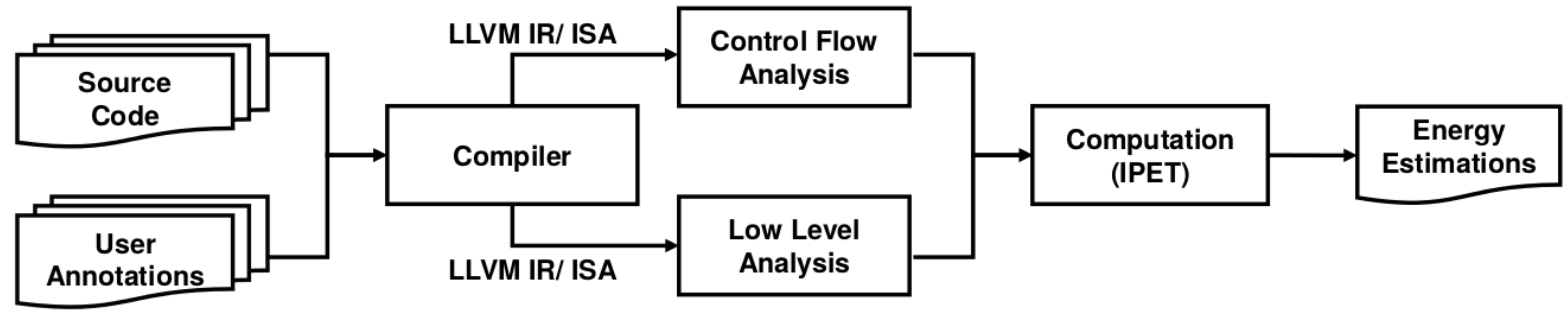}
\caption{Overview of our energy consumption static analysis.}
\label{fig:SAoverview}
\end{figure*}

\Cref{fig:SAoverview} shows the ECSA process for both, analysis at ISA and LLVM IR level. The source code together with any user annotations (e.g.\ to provide loop bounds) is sent to the compiler which emits the LLVM IR and the ISA code. Low level analysis, analysis of program control flow and computation of the energy consumption estimations is then applied on both levels. For the LLVM IR analysis an extra step is required at the compilation phase for the energy characterization of the LLVM IR instructions as detailed in \Cref{subsec:mapping}. In the rest of this section we briefly introduce each ECSA stage.

\subsection{Low Level Analysis}
\label{subsec:LowLevelAnalysis}

This stage aims to model the micro-architecture dynamic behavior of the processor based on an ISA-level energy model.

\subsubsection{XMOS Xcore ISA level Energy Modeling}
\label{sec:xs1model}

The Xcore processor is hardware multi-threaded, providing inter-thread
communication and I/O port control directly in the ISA. It is event-driven; busy waiting is avoided in favor of hardware scheduled idle periods. This makes the Xcore well suited to embedded applications requiring multiple hardware interfaces with real-time responsiveness. 

The underlying energy model for this work is captured at the ISA level.
Individual instructions from the ISA are assigned a single cost each. These can
then be used to compute power or energy for sequences of instructions. The
model also captures the cost of thread scheduling performed by the hardware, in
accordance with a series of profiling tests and measurements, because it
influences the energy consumption of program execution. Instructions from
runnable threads are scheduled round-robin by the hardware. To avoid data
hazards, the processor's four stage pipeline may only contain one instruction
from each thread. If the number of runnable threads is less than four, there
will be empty pipeline stages.

The modeling technique is built upon~\cite{Tiwari1996}, as discussed in \Cref{sec:modeling_review}, which is adapted and extended to consider the scheduling behavior and pipeline characteristics of the Xcore~\cite{Kerrison15}. 
A new version of this model that is well suited for static analysis has been developed. It represents energy in terms of static and dynamic power components to better reflect inter-instruction and inter-thread overheads. This has improved model accuracy by an average of 4\%.

\begin{equation}
  E_\text{prg} =  \left(P_s + P_{di}\right) \cdot T_\text{idl} + \sum_{i \in
        \text{prg}}\left(
          \frac{P_s + P_i M_{N_p} O}{N_p} \cdot 4 \cdot T_\text{clk}
        \right)
\text{, where } N_p = \min(N_t, 4)
   \label{eq:xs1model_new_mt}
\end{equation}

In \Cref{eq:xs1model_new_mt}, $E_\text{prg}$ is the energy of a program, formed
by adding  the energy consumed at idle to the energy consumed by every
instruction, $i$, executed in the program. At idle, only a base processor
power, the sum of its static, $P_s$, and dynamic idle power, $P_\text{di}$, is
dissipated for the total idle time, $T_\text{idl}$. For each instruction,
static power is again considered, with additional dynamic power for each
particular instruction, $P_i$. The dynamic power contribution is then
multiplied by a constant inter-instruction overhead, $O$, that has been
established as the average overhead of instruction interleaving. This is then
multiplied by a scaling factor to account for the number of threads in the
pipeline, $M_{N_p}$. The result is divided by the number of instructions in the
pipeline, which is at most four and is dependent upon the number of active
threads, $N_t$. Each instruction completes in four cycles, so $4 \cdot
T_\text{clk}$ gives the energy contribution of the given instruction, based on
the calculated power.

When more than four threads are active, the issue rate of
instructions per thread will be reduced. The energy model accounts for this with
the $\min$ term in \Cref{eq:xs1model_new_mt}. From a purely timing perspective,
the latency between instruction issues for a thread is $\max\left(N_t,4\right)
\cdot T_\text{clk}$. This property means that instructions are
time-deterministic, provided the number of active threads is known. A thread
may stall in order to fetch the next instruction. This is also deterministic
and can be statically identified~\cite[pp.\  8--10]{xs1architecture}. These
instruction timing rules have been used in simulation based energy estimation,
and are also utilized in the multi-threaded static analysis performed in this
paper. Both simulation and static analysis must be able to determine $N_t$, the
number of active threads, in order to accurately estimate energy consumption.

A limited number of instructions can be exceptions to these timing rules. The divide and remainder instructions are bit-serial and take up to 32 cycles to complete. Resource instructions may block if a condition of their execution is not met, e.g.\ waiting on inbound communication causes the instruction's thread to be de-scheduled until the condition becomes satisfied. This paper focuses its contributions on fully predictable instructions, with timing disturbances from communication forming future work.

The cost associated with an instruction represents the average energy consumption obtained from measuring the energy consumed during instruction execution based on constrained pseudo-randomly generated operands using the setup described in~\cite{Kerrison15}.
Thus, this model does not explicitly consider the range of input data values and how this may affect consumption. Empirical evidence indicates that such factors can contribute to the dynamic energy consumption~\cite{DBLP:journals/corr/PallisterKME15}.
The implications of using a random data constructed single value energy model with a bound SRA for ECSA are discussed in \Cref{sec:results}.

\subsubsection{Utilizing the Xcore energy model in static analysis}
\label{subsub:utilizingEM}

To determine the energy consumption of a program based on \Cref{eq:xs1model_new_mt} the  program's instruction sequence, $\langle i_1, \dots, i_n\rangle$, the idle time $T_\text{idl}$, and the number of active threads $N_p$ during instruction execution must be known.
In~\cite{Kerrison15} Instruction Set Simulation (ISS) was used to gather full trace
data or execution statistics to obtain these parameters. In this work we use ISS only as a reference for comparison of ECSA results, with a second reference being direct hardware measurement. 

ECSA thus needs to extract the CFGs for each thread and identify the interleavings between them. This allows for each instruction in the program to identify the $N_p$ component in \Cref{eq:xs1model_new_mt}. It also allows to estimate the total idle time, $T_\text{idl}$
of the program. 
For single-threaded programs the energy characterization of the CFG is straightforward as there is no thread interleaving. The IPET can be directly applied on the energy characterized CFG to extract a path that bounds the energy consumption of the program, as described in \Cref{subsec:ipe}. For arbitrary multi-threaded programs, energy characterizing the CFGs of each thread using static analysis is challenging. We have therefore concentrated on two commonly used concurrency patterns, task farms and pipelines, which we use with evenly distributed workloads across threads. 

In addition to the instructions defined in the ISA, a Fetch No-Op (FNOP) can also be issued by the processor. These occur deterministically~\cite[pp.\ 8--10]{xs1architecture}. FNOPs can have a significant impact on energy consumption, particularly within loops. To account for FNOPs in static
analysis, the program's CFG at ISA level is analyzed. An instruction buffer model is used to determine where FNOPs will occur in a basic block. However, one particular FNOP case is dependent on the dynamic branching behavior of the program, in which case we over-estimate FNOPs. Further implementation details on FNOPs modeling can be found in~\cite{fnops}.

\subsection{Mapping an ISA Energy Model to LLVM IR}
\label{subsec:mapping}

Although substantial effort has been devoted to ISA energy modeling, there is little research into modeling at higher levels of program representation, where precision can decrease. In~\cite{Brandolese2011}, statistical analysis and characterization of LLVM IR code is performed. This is combined with instrumentation and execution on a target host machine to estimate the performance and energy requirements in embedded software. Transferring the LLVM IR energy model to a new platform requires performing the statistical analysis again. The mapping technique we present here is fully portable. It requires only the adjustment of the LLVM mapping pass to the new architecture. Furthermore, our LLVM IR mapping technique provides on-the-fly energy characterization that allows to take into consideration the compiler behavior, CFG structure, types and other aspects of instructions.

\subsubsection{Formal specification of the mapping}
\label{mappingFormal}

Our mapping technique determines the energy characteristics of LLVM IR
instructions. Thus, mapping links LLVM IR instructions with machine specific
ISA instructions. ISA level energy models can then be propagated up to LLVM IR
level, allowing energy consumption estimation of programs at that level.
We formalize the mapping as follows. For a program $P$, let
\begin{equation}
    \mathrm{IRprog}_L = \{1,2,...,n\}
\label{eq:LLVMIDs}
\end{equation}
be the ID numbers of $P$'s LLVM IR instructions and therefore
\begin{equation}
  \mathrm{IRprog}=\langle ir_{1},ir_{2},...,ir_{n}\rangle
\label{eq:LLVMIRcode}
\end{equation}
is the sequence of LLVM IR instructions for $P$.
\begin{equation}
 \mathrm{T}_{arch}(\mathrm{IRprog}) = \mathrm{ISAprog}
\end{equation}
is an architecture specific compiler back end that can translate the IRprog to 
\begin{equation}
  \mathrm{ISAprog}=\langle (isa_{1},m_1),(isa_{2},m_2),...,(isa_{k},m_l) \rangle \text{ where } m_1, m_2,...,m_l \in \mathrm{IRprog}_L
  \label{eq:ISAcode}
\end{equation}
which is the sequence of ISA instructions for $P$, together with the ID of the LLVMIR instruction from which each $isa_k$ originated. If an $isa_k$ comes from more than one LLVM IR instructions, then $\mathrm{T}_{arch}$ chooses the ID of one of them to assign to $isa_k$.

\begin{equation}
\begin{split}
& \mathrm{M}(ir_i) = \{ isa_j | ir_i \in \mathrm{IRprog} \wedge \mathrm{ISAprog} = \mathrm{T}_{arch}(\mathrm{IRprog}) \wedge (isa_{j},i) \in \mathrm{ISAprog} \}  \text{ and }\\
& \forall ~ ir_n, ir_k \in \mathrm{IRprog} \wedge ir_{n} \neq ir_k \text{ then }
 \mathrm{M}(ir_n) \cap \mathrm{M}(ir_k) = \emptyset
\end{split}
\label{eq:mapping}
\end{equation}
is a mapping function that captures a 1:m relation from $\mathrm{IRprog}$ to $\mathrm{ISAprog}$ instructions. Therefore,
 \begin{equation}
   \mathrm{E}(ir_{i}) = \sum_{isa_j \in S} \mathrm{E}(isa_{j}) \text{ where } ir_{i} \in \mathrm{IRprog} \wedge isa_{j} \in \mathrm{ISAprog} \wedge \mathrm{S}=\mathrm{M}(ir_i)
 \label{eq:energy}
\end{equation}
represents the energy consumption of an LLVM IR instruction as the sum of the energy consumed by all ISA instructions associated with that LLVM IR instruction.

By instantiating the above mapping to a specific architecture, LLVM IR energy characterization can be retrieved. The accuracy of this characterization can vary for different architectures. If the accuracy is not adequate, then a tunning phase can be introduced to account for any specific compiler or architecture behavior. An example of such tunning is given in the next section, which accounts for \emph{phi-nodes} and FNOPs.

\subsubsection{Xcore mapping instantiation and tuning}

In our case, the $\mathrm{T}_{arch}$ function is the XMOS tool chain lowering phase that translates the LLVM IR to Xcore specific ISA. Our mapping implementation leverages the debug mechanism in the XMOS compiler tool chain, in order to enable $\mathrm{T}_{arch}$ to assign to each ISA instruction the ID of the LLVM IR instruction it originated from. This is typically used by the programmer to identify and fix problems in application code. Debug symbols are created during compilation to assist with this. These symbols are propagated to all intermediate code layers and down to the ISA code. Debug symbols can express which programming language constructs generated a specific piece of machine code in a given executable module. In our case, these symbols are generated by the front end of the XMOS compiler in standard DWARF format~\cite{DWARF:2013:Online}. These are transformed to LLVM metadata~\cite{LLVMmetaData:2014:Online} and attached to the LLVM IR.

During the lowering phase of compilation, LLVM IR code is transformed to a target ISA by the back end of the compiler, with debug information stored alongside it as LLVM metadata. Naturally, the accuracy of debug information in the output executable is reduced if the number of optimization passes is increased. This is due to portions of the initial LLVM IR either being discarded or merged during these passes.

Tracking this information gives an $n:m$ relationship between instructions at the different layers, because source code instructions can be translated to many LLVM IR instructions, and these again into many ISA instructions. This $n:m$ relation prevents ECSA from providing accurate energy values and therefore the mapping introduced in \Cref{mappingFormal}, requires~\Cref{eq:mapping} to create an 1:m relation between the LLVM IR and ISA code.

To address this issue, we created an LLVM pass that traverses the LLVM IR and replaces source location information with LLVM IR location information. The location information represents the $\mathrm{IRprog}_L$ in \Cref{eq:LLVMIDs}. The LLVM pass runs after all optimization passes, just before emitting ISA code. The optimized LLVM IR is closer in structure to the ISA code than the unoptimized version. Using this method a $1:m$ mapping between LLVM IR instructions and ISA instructions can be extracted by \Cref{eq:mapping}. Once the mapping has been performed for a program, the energy values for groups of ISA instructions are aggregated and then associated with their single corresponding LLVM IR instruction using \Cref{eq:energy}.

\begin{figure}
        \centering
        \includegraphics[width=1\textwidth,clip,trim=0.38cm 7.5cm 1cm 3cm]{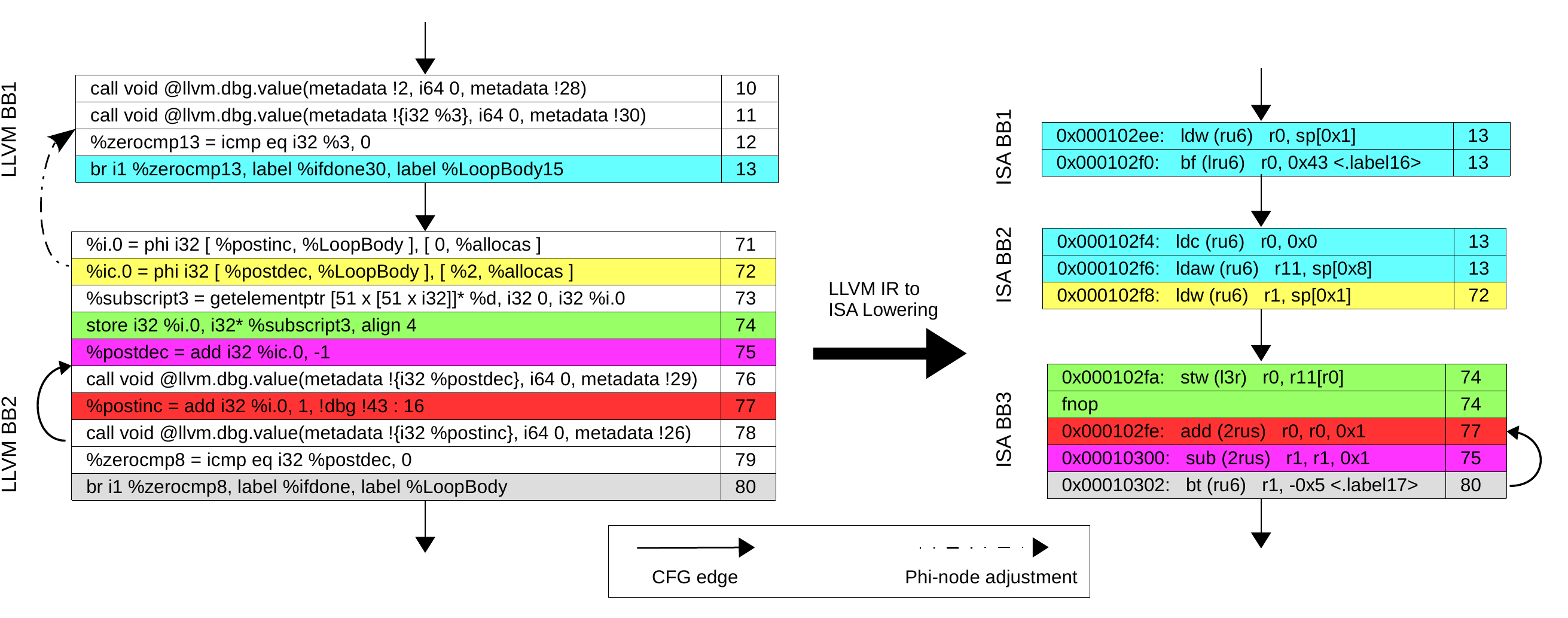}%
        \caption{Fine grained 1:m mapping including our LLVM mapping pass.}
        \label{fig:mapping}
\end{figure}

An example mapping is given in \Cref{fig:mapping}. On the left hand side is a part of the LLVM IR CFG of a program, which represents the $\mathrm{IRprog}$ in \Cref{eq:LLVMIRcode}, along with the debug location, $\mathrm{LLVMIR}_L$ in \Cref{eq:LLVMIDs}, for each LLVM IR instruction. The right hand side shows the corresponding ISA CFG, together with the debug locations for each ISA instruction, given by $\mathrm{T}_{arch}$. The coloring of the instructions demonstrates the mapping between the two CFGs' instructions using \Cref{eq:mapping}. Now, one LLVM IR instruction is matched to many ISA instructions, but each ISA instruction is mapped to only one LLVM IR instruction. Some LLVM IR instructions are not mapped, because they are removed during the lowering phase of the compiler. This mapping also guarantees that all ISA instructions are mapped to LLVM IR, so there is no loss of recorded energy between the two levels.

Additional optimizations are performed during the lowering phase from LLVM IR to ISA, such as peephole optimizations and target specific optimizations. These can affect the mapping, but not to the same degree as the LLVM optimizations. A tuning phase can be introduced after the mapping, to account for them.

The mapping instantiation for the Xcore architecture was able to provide an average energy estimation deviation of 6\% from the predictions on the ISA level. An additional tuning phase is introduced after the mapping, to account for specific compiler and architecture behavior. This improved the mapping accuracy, narrowing the gap between ISA and LLVM IR energy predictions to an average of 1\% as discussed in~\ref{LLVMIRuse}.

LLVM IR \emph{phi-nodes} are an example of such tuning. \emph{Phi-nodes} can be introduced at the start of a BB as a side effect of the Single Static Assignment (SSA) used for variables in the LLVM IR. A \emph{phi-node} takes a list of pairs, where each pair contains a reference to the predecessor block together with the variable that is propagated from there to the current block.  The number of pairs is equal to the number of predecessor blocks to the current block. A \emph{phi-node} can create inaccuracies in the mapping when LLVM IR is lowered to ISA code that no longer supports SSA, because it can be hoisted out from its current block to the corresponding predecessor block. For blocks in loops this can lead to a significant analysis error.

Whenever the tuning phase is able to track these cases, it can adjust the energy figures for each LLVM IR BB accordingly. An example of this is given in \Cref{fig:mapping} at debug location number 72. Its corresponding ISA instruction is hoisted out from the loop BB \texttt{ISA BB3} and into
\texttt{ISA BB2}. This is tracked by the mapping, and the equivalent hoisting is done at LLVM IR level, thus correctly assigning energy values to each LLVM IR block. Similar errors can be introduced by branching LLVM IR instructions with multiple targets, since in the Xcore ISA only single target branches are supported. This is also handled during the mapping phase.

As discussed in \Cref{subsub:utilizingEM}, FNOPs can be issued by the processor and this can be statically determined at the ISA level. LLVM IR has no way to represent this. Ignoring them can therefore lead to a significant underestimation of energy at LLVM IR level. To address this, FNOPs in the lowered ISA code are assigned the debug location of an adjacent ISA instruction in the same BB by the tuning phase, thus they are accounted for in the mapped LLVM IR block.

LLVM IR instructions can be combined into a single ISA instruction. An example of such instructions are the add and multiplication ones which can be translated to the Xcore \texttt{macc} (multiply-accumulate) ISA instruction. The  $\mathrm{T}_{arch}$ will assign the energy cost to only one of the LLVM IR instructions. Although, this is adequate for the energy characterization of LLVM IR basic blocks, if needed the tuning phase allows to associate the cost with both instructions.

\begin{figure*}[ht]
\centering
\includegraphics[width=1.0\textwidth]{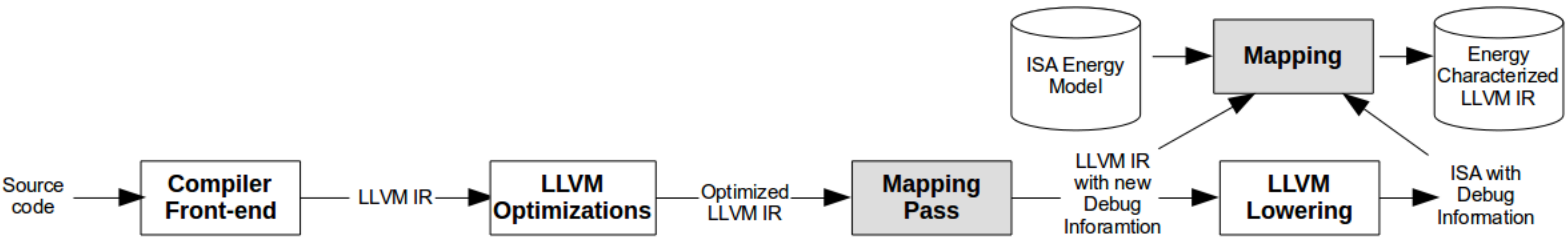}
\caption{Overview of the mapping process.}
\label{fig:Mappingoverview}
\end{figure*}

An overview of the mapping technique is given in \Cref{fig:Mappingoverview}. Our mapping pass is introduced into the compilation process after LLVM optimizations. The pass also includes tuning. The mapping phase implements Relations~\ref{eq:mapping} and~\ref{eq:energy}. It runs after the LLVM lowering phase and maps LLVM IR instructions with the new debug locations to the emitted ISA instructions. The ISA energy model is then used to accumulate the energy value of each LLVM IR instruction based on its mapped ISA instructions.

\subsection{Control Flow Analysis}

This component aims to capture the dynamic behavior of the program and
associates CFG BBs with the information needed for the computation step of
analysis. IPET requires the CFG and call graph of a program to be constructed
at the same level as the analysis. At LLVM IR level, the compiler can generate
them. At ISA level a tool was created to construct them. To detect BBs that
belong to a loop or recursion, we adopted and extended the algorithm
in~\cite{Wei:2007}. The CFGs are annotated according to the needs of the IPET
described in \Cref{subsec:ipe}. Finally, the annotated CFGs are used in the
computation step to produce ILP formulations and constraints.

\subsection{Computation}
\label{subsec:ipe}

The IPET adopted in our work to estimate the energy consumption of a program is
based on~\cite{Li1997}. To construct the ILP system needed for IPET, we use
information produced from the previous two components. The method of ILP
formulation along with the constraints needed to bound the problem and optimize
it's solution can be found in the seminal paper~\cite{Li1997}. To infer the
energy consumption, instead of using the time cost of a CFG basic block we are
using its energy cost, as provided by the respective energy model.

Constraints are used to capture information that can affect a program's dynamic
behavior, such as bounded loop iterations, or path information, such as
infeasible paths. Usually, this information can only be specified by the
programmer, as it depends on the program semantics and cannot be extracted by
the static analysis. The minimum required user input to enable
bounding of the problem is the declaration of loop bounds. This is also
standard practice in timing analysis~\cite{Wilhelm:2008}. Providing this kind
of information is usually easy, as the loop bounds are typically known by
programmers of timing critical embedded programs. Further constraints, such as
denoting infeasible paths in a CFG, can be provided to extract more accurate
estimations. The user provides this information as source code annotations. The
annotation language used in this work can be found in~\cite{entra-d2.1}.

\subsection{Analysis of multi-threaded programs}
\label{subsec:mult-threadedAnalysis}

In this paper we present the first steps towards ECSA of multi-threaded
programs. Two concurrency patterns are considered: replicated threads with no
inter-thread communication, working on different sets of data (task farms), and
pipelines of communicating threads. For both cases, we consider evenly
distributed, balanced work loads. In the former case, an example use is
simultaneously processing multiple independent data. In the latter case,
pipelining enables parallelism to be used to improve performance when
processing a single data stream.

There is a fundamental difference when statically predicting the case of
interest (worst, best, average case) for time and for energy for multi-threaded
programs. Generally, for time only the computations that contribute to the path
forming the case of interest must be considered. For energy, all computations
taking place during the case of interest must be considered. For instance, in
an unbalanced task farm, the WCET will be equivalent to the longest running
thread. To bound energy, the energy consumption of each thread needs to be
aggregated. This is harder since the static analysis needs to determine the number
of active threads at each point in time in order to apply the energy model from
\Cref{eq:xs1model_new_mt} and characterize the CFG of each thread. Then,
IPET can be applied to each thread's CFG, extracting energy
consumption bounds. Aggregating these together will give a loose upper bound on
the program's energy consumption, meaning that the safety of the bound cannot be guaranteed.

In our balanced task farm examples, all task threads are active in parallel for
the duration of the test. Thus, the number of active threads is constant,
giving a constant $N_t$, used to determine the pipeline occupancy scaling
factor, $M$, in \Cref{eq:xs1model_new_mt}. For balanced pipelined programs, we
consider the continuous, streaming data use case, so the same constant thread
count property holds. In both cases IPET can be performed on each thread's CFG
and the results aggregated to retrieve the total energy consumption. In this
work, core-local communication is considered, which uses the same instructions
as off-core communication, but no external link energy needs to be accounted
for. Therefore the core energy model provides sufficient data.

For multi-threaded programs with synchronous communications, to retrieve a
WCET, IPET can be applied on a global graph, connecting the CFGs of all threads
along communication edges. The communication edges can be treated by the IPET
as normal CFG edges and WCET can be extracted by solving the formulated
problem~\cite{PotopButucaru:2013}. This will return a single worst case path
across the global graph. Bounding energy in this way is not possible, as
parallel thread activity over time needs to be considered.  Here the task is
even harder in comparison to programs without communication, as activity can be
blocked if the threads' workloads are unbalanced. In this case, statically
determining the number of active threads at each point in time is a hard challenge.

Although the concurrency patterns addressed here can be considered as easy targets
for the ECSA, they are typical embedded use cases, and as is explained in
\Cref{multi-threaded_usage}, ECSA can provide sufficiently accurate information to enable
energy aware decision making. Building on this, more complex programs will be
analyzed in future work, such as unique non-communicating threads rather than
replicated threads, unbalanced farm and pipeline workloads and other concurrency patterns. Such programs will feature varying numbers of active threads over the course of execution. In these cases the ECSA must be extended to perform analysis that extracts all the possible combinations of thread interleaving.

This work focuses on multi-threaded communication on a single core. However,
for communicating threads, the channel communication paradigms that are used by
the programs at the source code level and within the ISA can also be used in a
multi-core environment, creating scope for the analysis of larger systems.

\section{Experimental Evaluation}
\label{sec:evaluation}

To evaluate our ECSA, a series of mainly industrial benchmarks were selected with representative test cases. Both our ECSA results and estimations from ISS using the same energy model are compared to hardware measurements. The benchmarks, evaluation methodology, results and further observations are discussed in this section.

\subsection{Benchmarks}

Our objective is to demonstrate the value of our ECSA for common industrial, deeply embedded applications. A complete list of all the 21 benchmarks' code and summary of their attributes, can be found in~\cite{benchmarks}. Benchmarks were compiled with \texttt{xcc} version 12~\cite{xtools} at optimization level \texttt{O2}; the default for most compilers.

Deeply embedded processors do not typically have hardware support for division
or floating point operations, using software libraries instead. Software
implementations are usually far less efficient than their hardware equivalent,
both in terms of execution time and energy consumption. The effect of these software implementations on energy consumption should be known by developers, therefore we include soft division and soft float benchmarks.

A radix-4 software divider, \texttt{Radix4Div}~\cite{Radix4Div}, is used. A
less efficient version, \texttt{B.Radix4Div}, is added for comparison. This
version omits an early return when the dividend is greater than 255. A
consequence of excluding this optimization is that CFG paths become more
balanced, with less variation between the possible execution paths. The
effect of this on the energy consumption is discussed later in this section. For software floating point, single precision \texttt{SFloatAdd32bit} and \texttt{SFloatSub32bit} operations from~\cite{SoftFloat} are analyzed.

To represent common signal processing tasks, \texttt{FIR} and \texttt{Biquad}
benchmarks written for the Xcore processor~\cite{XMOSDSP}, are analyzed. In
addition, a series of open source benchmarks of core algorithmic functions were
selected from the \texttt{MDH WCET} benchmark suite~\cite{Gustafsson2010}. They
were modified to work with our test harness and, in some cases, to make them
more parametric to function input arguments. Some were extended to be
multi-threaded task farms, where the same code runs on two or four threads.  To
extend our analysis to multi-threaded communication programs, we analyze
pipelined versions of \texttt{FIR} and \texttt{Biquad}, each formed of seven
threads. These programs are the preferred form for Xcore, as spreading the
computation across threads allows the voltage and frequency of the core to be
lowered, significantly reducing energy consumption with the same performance as
the single threaded version.

\subsection{Results Analysis}
\label{sec:results}

\begin{figure}[!ht]
    \begin{minipage}{1.0\textwidth}
  \centering
  \includegraphicsmaybe{width=1.0\textwidth}{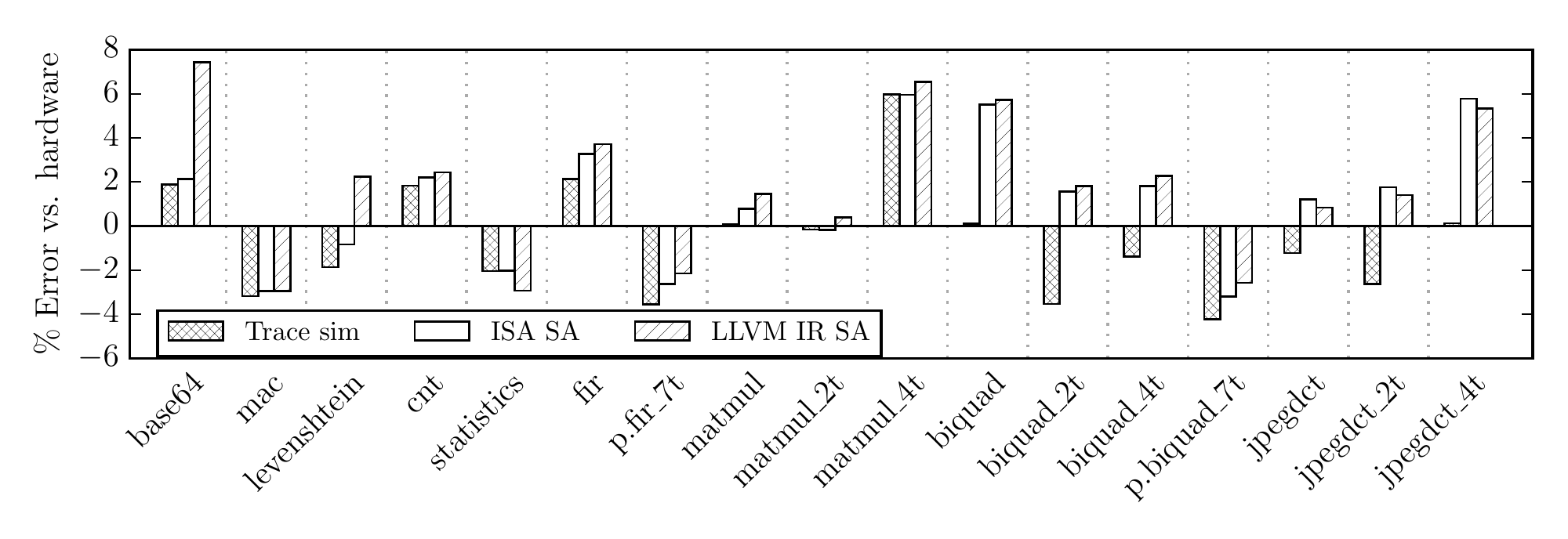}
  \vspace{-8mm}

  \subcaption{\label{fig:results1:a}All benchmarks.}
    \end{minipage}
    \begin{minipage}{1.0\textwidth}
        \centering
        \includegraphicsmaybe{width=0.32\textwidth,trim=0.7cm 0cm 0cm 0cm,clip}{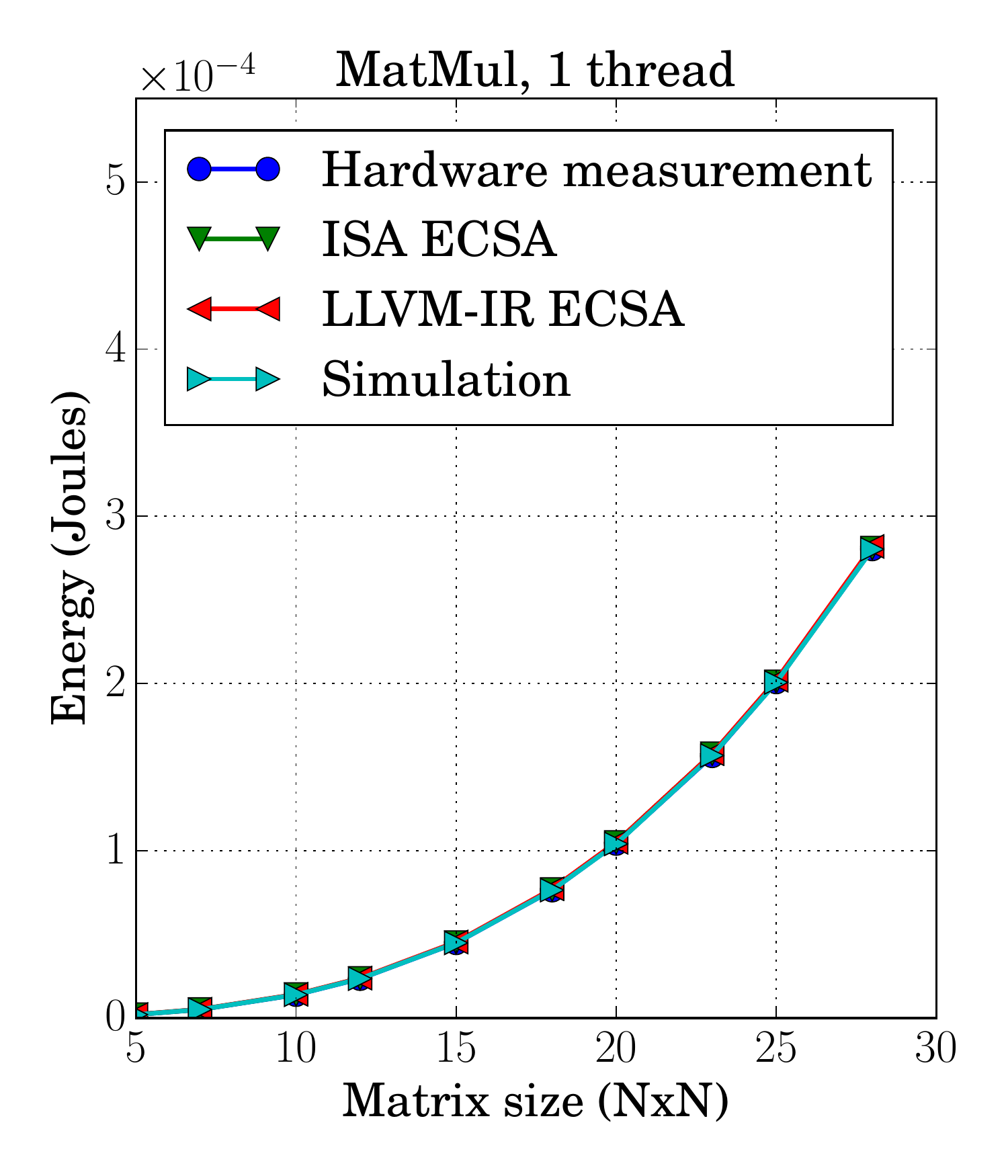}
        \includegraphicsmaybe{width=0.32\textwidth,trim=0.7cm 0cm 0cm 0cm,clip}{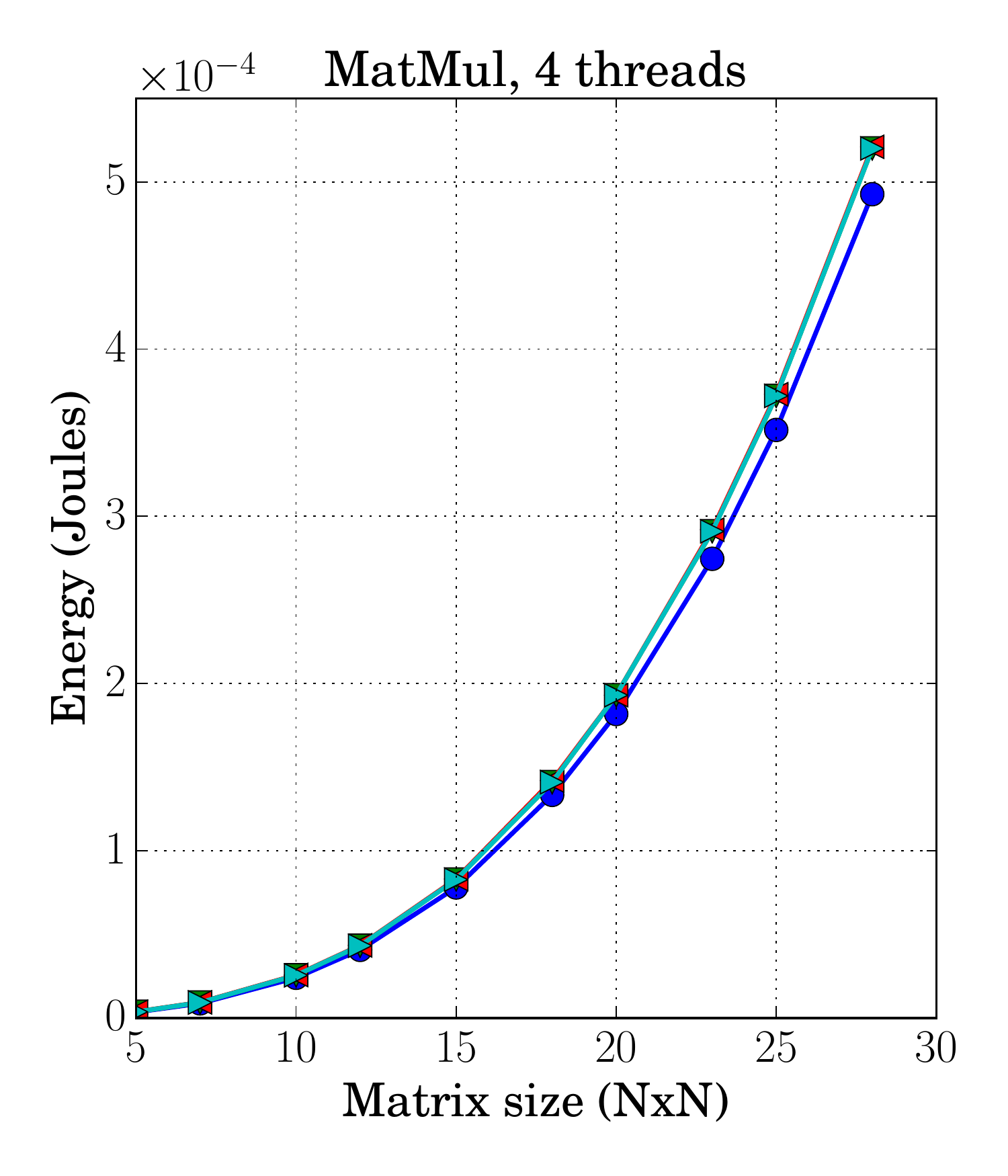}
        \includegraphicsmaybe{width=0.32\textwidth,trim=0.75cm 0cm 0cm 0cm,clip}{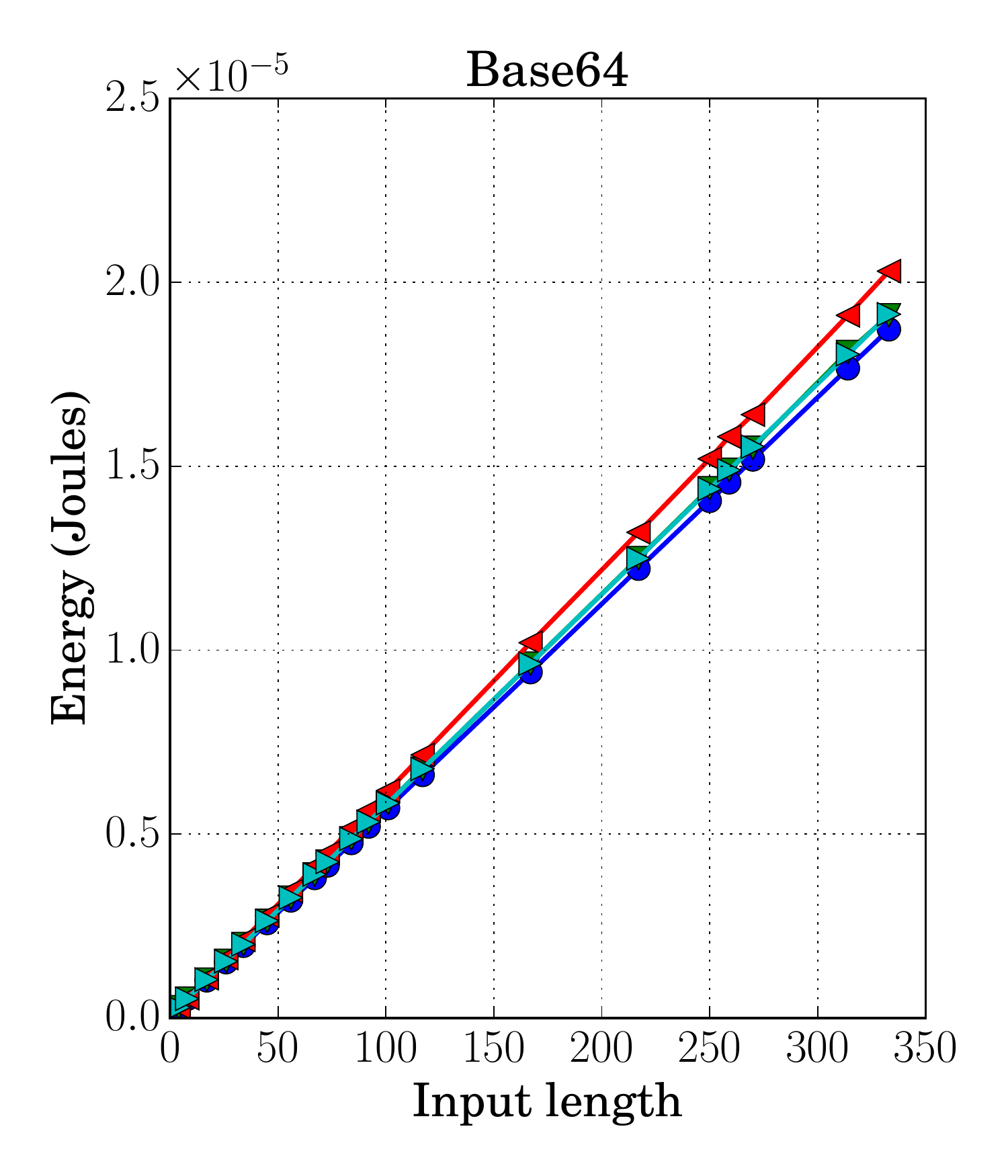}
    \subcaption{\label{fig:results1:b}Parametric benchmarks.}
     \end{minipage}
    \caption{Hardware measurements compared to ECSA and ISA trace estimation.}
    \label{fig:results1}
\end{figure}

The experimental results show several features, influenced by the level of
multi-threading, the properties of the benchmarks, and the 
levels at which ECSA and modeling are performed. In this section we examine
all of these in order to determine what influences ECSA accuracy at each
level, highlighting both strengths and limitations. 

\Cref{fig:results1:a} presents the error margin of using our energy model with
three energy estimation techniques compared to hardware energy measurements for our
benchmarks. \emph{Trace Sim} produces instruction traces from ISS,
\emph{ISA ECSA} uses the model for static analysis at the ISA level and
\emph{LLVM IR ECSA} uses our mapping technique to apply the model and analysis
at LLVM IR level. For all benchmarks with multiple test parameters, the
geometric mean of the errors is used. \Cref{fig:results1:b} compares energy
estimates to hardware measurements for a range of parameters in three
parametric benchmarks.

For \texttt{Levenshtein}, \texttt{MatMult 1,2,4}, \texttt{Mac}, \texttt{Cnt}
and \texttt{Base64} parametric energy consumption estimations can be
determined, as discussed in \Cref{costFunctions}. These are expressed in terms of a function over the number of loop iterations.

The parametric benchmarks are also more data sensitive, due to the use of matrices. The hardware energy measurements for all the benchmarks using matrices were obtained by using random data to initialize them. In order to investigate the effect of different random data, the measurements were repeated 500 times for each benchmark using a different seed each time for data generation. The maximum variation observed was in the range of the measurement error, less than 0.5\%, and therefore the average of these measurements was used to compare against the predicted results. The effect of using non random data will be investigated in \Cref{dataEffect}. For the more industry oriented benchmarks (all the \texttt{FIR}, \texttt{Biquad} and \texttt{Jpegdct} versions) real sample data where used for the hardware measurements.

\begin{figure}[!ht]
    \centering
    \includegraphicsmaybe{width=0.44\textwidth,trim=0.75cm 0cm 0.70cm 0cm,clip}{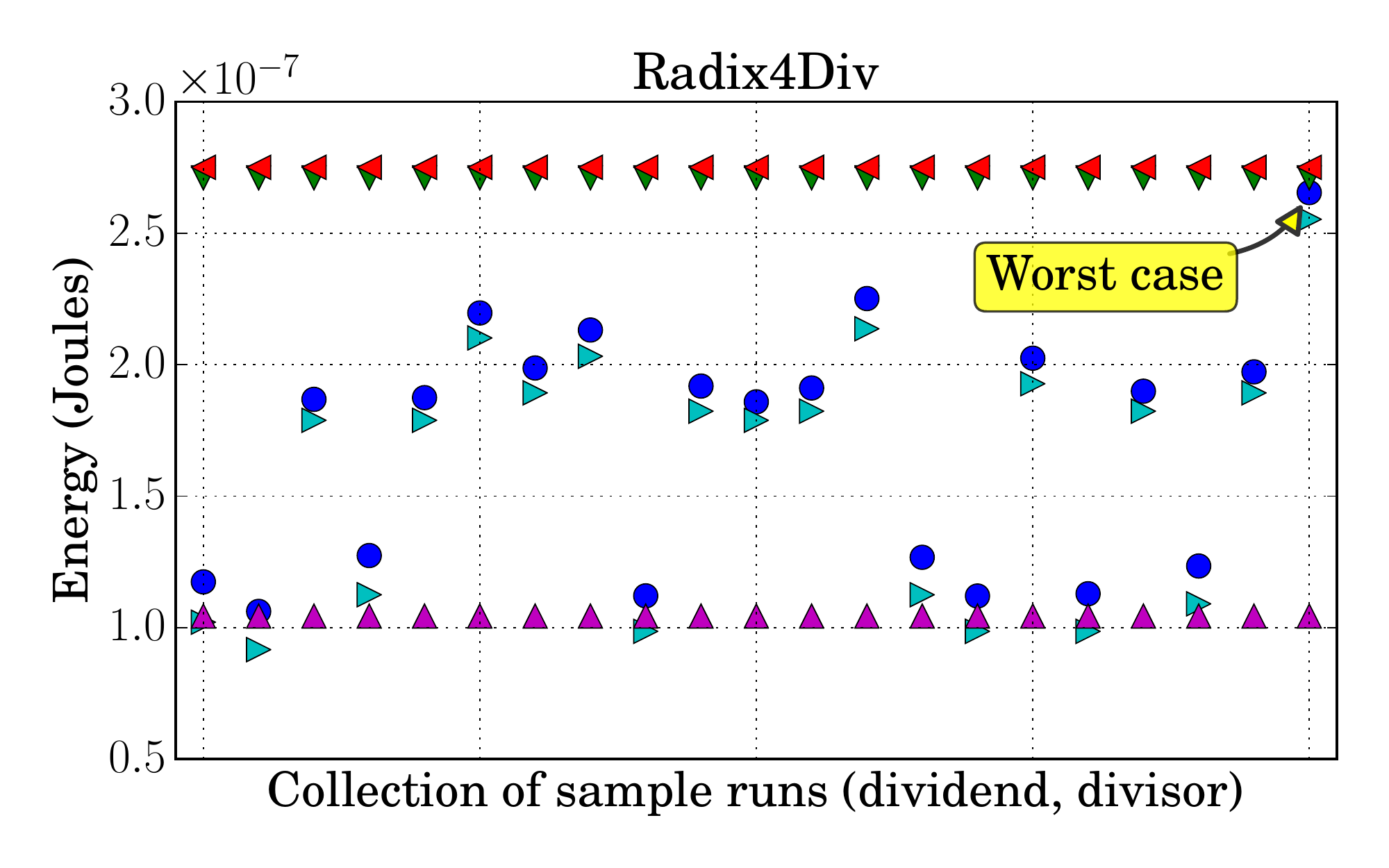}
    \includegraphicsmaybe{width=0.44\textwidth,trim=0.75cm 0cm 0.70cm 0cm,clip}{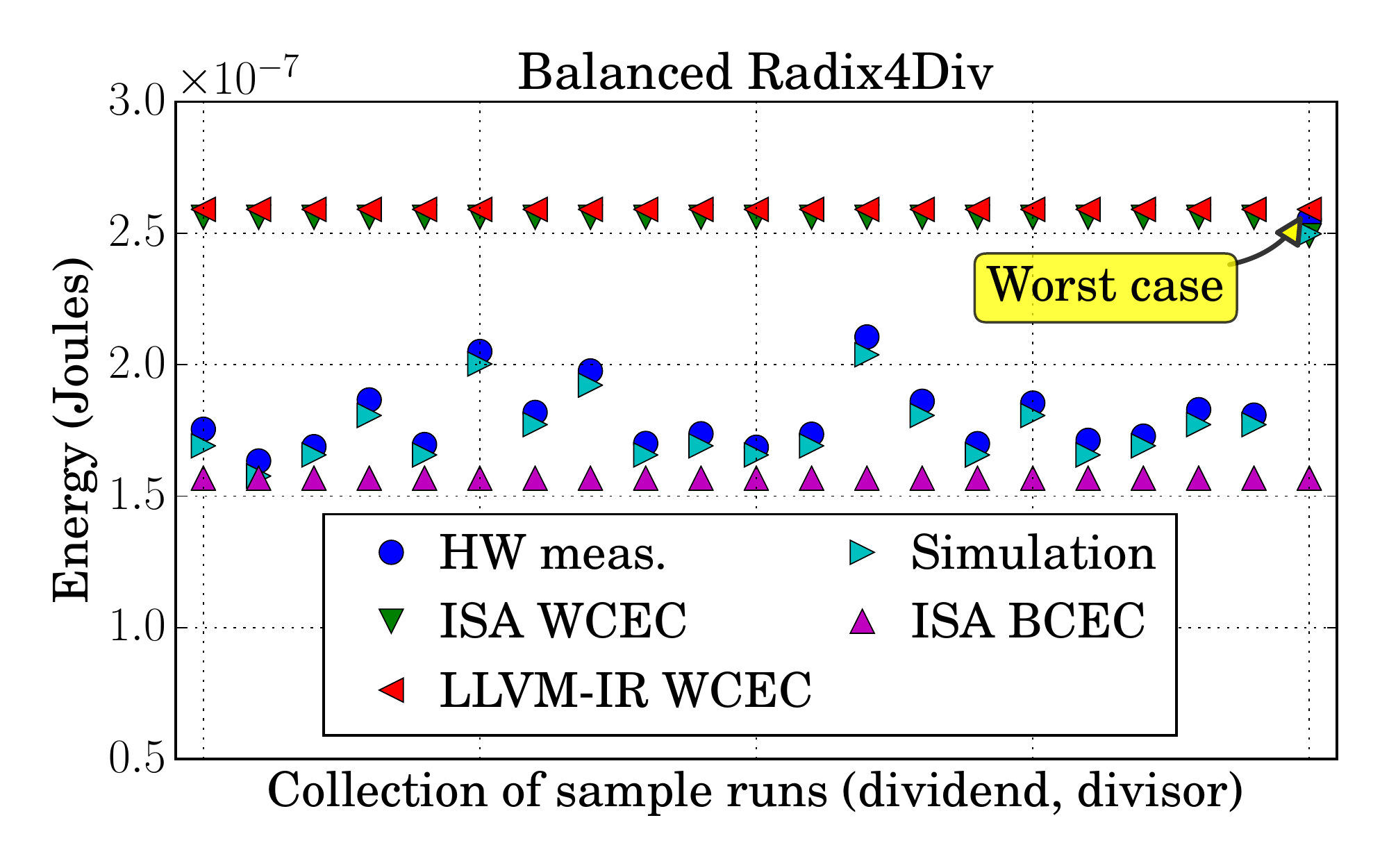}
    \caption{Results for benchmarks with constant ECSA estimations across all test cases.}
    \label{fig:results2}
\end{figure}

For the software division and floating point benchmarks, ECSA provides a constant energy consumption upper bound across all test cases, as they contain no loops that are directly affected by the functions' arguments. \Cref{fig:results2} demonstrates this for \texttt{Radix4Div} and \texttt{B.Radix4Div}. Considering that IPET is intended to provide bounds based on a given cost
model, in our case it tries to select the worst case execution paths in terms of the energy consumption. Therefore, the ECSA estimations seen in \Cref{fig:results2} represent a loose upper bound on the benchmarks' energy consumption. Similar figures were also retrieved for the two
\texttt{SoftFloat} benchmarks. These bounds, in most cases cannot be considered safe, as they might be undermined by the use of a non data sensitive energy model and analysis. However, they can still give the application programmer valuable guidance towards energy aware software development, as discussed in \Cref{boundUse}.

For the benchmarks in \Cref{fig:results2}, we sought test cases that exercise the average, best- and worst-case scenarios of each benchmark's algorithm, to compare the resultant range of energy consumption with our ECSA predictions. A good understanding of the underlying algorithms and information collected from the ISS traces was necessary to identify tests covering each scenario with certainty. This poses a challenge in guaranteeing that the cases of interest, such as worst case, have been exercised. For example, the \texttt{Radix4Div} benchmark takes two 16-bit parameters, forming a search-space of $2^{32}$ test cases. This was suitably small to perform an exhaustive search in order to capture the worst case empirically. However, the time cost of an exhaustive search precludes doing the same for many other benchmarks. For both \texttt{Radix4Div} variants, the upper bounds inferred by the IPET analysis are not only very close to the worst
case retrieved by exhaustively searching the possible test cases, but are also safe.

Generally, for all results shown, a proportion of error is present in both forms of static analysis as well as simulation based energy estimation. The error in the ISS based estimation is a baseline for the best achievable error in static analysis, as ISS produces more accurate execution information. For all the benchmarks, the ISA ECSA results are over-approximating the trace based energy estimations. This applies also to the LLVM IR ECSA results with exception of the \texttt{statistics} benchmark. This over-approximation is a product of the bound analysis used which is trying to select the most energy costly CFG path based on the provided cost model. A smaller difference between the ECSA results and the trace
based energy estimations indicates that the execution path selected by the IPET fits better the actual execution path of a benchmark.

\subsubsection{Measurement error analysis}

To assess the accuracy of ECSA predictions, reliable hardware measurements are required. 
We use a shunt resistor current sense circuit and data
sampling hardware to obtain power dissipation with sub-milliwatt accuracy. The
data capture process is explained in more detail in~\cite{Kerrison15}.

Measurements are subject to errors introduced through environmental factors. In
particular, temperature and electrical noise can result in variations
of the measured energy consumption for multiple runs of the same test. To
measure the effect of these factors on our platform, we executed the
\texttt{MatMult 4} thread benchmark 100,000 times. This benchmark was selected
because it is particularly power intensive and likely to affect the device
temperature the most. The variation observed on our hardware was less than
0.7\% which we consider negligible and close to the error margin of our
measurement equipment. These factors could have a more significant impact on
other platforms. It is therefore important to examine them when performing ECSA.

The test harness introduces a small error by repeatedly calling the benchmark
function within a loop. This is necessary to ensure an adequate number of power samples are taken during the test. However, the loop surrounding the call to the benchmark,
together with the function call itself, introduce an overhead. This overhead
can be significant, especially when the amount of computation in the loop body
is low. To mitigate this overhead, we ensure that the loop is as efficient as
possible and each benchmark sufficiently large in size.  Finally, measurements
were taken several times to ensure that results obtained were consistent, with
less than 0.5\% variation.

\subsubsection{LLVM IR analysis accuracy}

This form of analysis is solely dependent on the accuracy of the mapping
techniques presented in \Cref{subsec:mapping}. As shown in
\Cref{fig:results1:a,fig:results2}, for all benchmarks the LLVM IR ECSA results
are within one percentage point error of ISA ECSA results, except for the
\texttt{Base64} benchmark with a further 5.3 percentage points error. In this
case the CFGs of the two levels were significantly different due to BBs introduced from branches in the ISA level CFG. This is one of the few cases where the mapper was unable to accurately track the differences between the two CFGs.

\subsubsection{Multi-threading accuracy}

Three benchmarks, \texttt{MatMul}, \texttt{Biquad} and \texttt{JpegDCT}, were extended to multi-threaded versions, where each thread executes the same program and processes its own data stream. The computation performed and the energy consumption increases with the number of active threads, with a negligible change in execution time. The underlying energy model is parametric to the occupancy of the execution pipeline, which is determined by the number of running threads. As such, the estimations from the model and their relative errors can differ when the number of threads is changed. For any given number of threads, the accuracy of the ECSA is influenced by the accuracy of the ISA level energy model. 

In the case of pipelined benchmarks, \texttt{p.fir\_7t} and
\texttt{p.biquad\_7t}, the energy model underestimates energy consumption by
approximately 5\%. This error is inherited by the ECSA. Further calibration of
the model is required to achieve better accuracy for multi-threaded programs
with communications.

\subsubsection{Data effect}
\label{dataEffect}

Since a non data sensitive ECSA will provide a single energy estimation regardless of input data values, comparing this to hardware measurements may give a different error for each input data. To examine this, we used one of our most data sensitive benchmarks, \texttt{MatMul 4}. 
The smallest over-estimation when compared with the hardware measurements was 5.96\% for both the ISA ECSA and trace based energy estimations. 
This was obtained for matrices that were initialized with randomly generated data.
Since our energy model was characterized using pseudo-randomly generated data, it provides a good fit to the data used for measurement, thus this result meets our expectations.
The maximum over-estimation found was 25\%, by initiating both matrices with zero data, minimizing the processor's switching activity. 
By using the same random data in the two matrices, the over-estimation was between the two previous cases, approximately 15\%. This is because the processor switching activity is less than in the case of different random data initialized matrices, and more than the case with the zero initialized matrices. Thus, users must be cautious when using ECSA with data sensitive benchmarks, as we will discuss in \Cref{ECSAapplications}.

These findings lead to two new research questions. Firstly, for convenience, many energy models are constructed from random input data. However, as we demonstrated, the closer the data used to characterize the energy model fits the data of the use case, the more accurate the ECSA estimation.
For example, \texttt{MatMul} and \texttt{fdct} are heavily used in video processing applications with highly correlated data between frames in the video stream. Therefore, a random data constructed energy model for these applications may not be suitable. How can we construct energy models that are more fit for purpose? Secondly, if a data sensitive energy model were to be constructed, how would this model be composed to be useful for ECSA? These two research questions motivate future work in this area.

\subsubsection{Static analysis limitations}

ECSA suffers from all the static analysis limitations that the timing analysis faces~\cite{Wilhelm2014}. Many of the techniques used by the timing analysis community to tackle these limitations can also be adopted in ECSA. For example, infeasible paths can lead to unrealistic estimations in both cases, energy and time. Techniques such as symbolic execution~\cite{wagemannworst} used in timing analysis to exclude infeasible paths, can be also used for ECSA. For this paper, source code annotations were translated to ILP constraints, in order to exclude infeasible paths from ECSA.

As already identified, ECSA can be more complicated than timing analysis. In \Cref{dataEffect}, we discussed that energy consumption is sensitive to the data related switching activity in the processor, which time is not affected from. In \Cref{subsec:mult-threadedAnalysis} we discussed, that for multi-threaded programs, timing analysis is considered only with a single path across all the threads, but ECSA has to consider all computations active during the case of interest.

In summary, the results show that static analysis, both at ISA and LLVM-IR
level, can deliver practical energy consumption estimates for a good range of
single and multi-threaded programs. The estimation error for both static and simulation based techniques can be reduced if the accuracy of the underlying energy model is improved.

\subsection{ECSA applications}
\label{ECSAapplications}

Precise energy measurements are often not easily accessible, requiring extra
equipment and hardware knowledge as well as modifications to the target
hardware. This makes it very difficult for most programmers to
assess a program's energy consumption. ECSA overcomes these obstacles by
providing energy transparency to users and systems with a useful level of
accuracy.

Trace based energy estimation allows for a very precise estimation of energy
consumption for a particular program run. The program is executed in simulation
with a given set of input parameters. The exact sequence of instructions can be
recorded during simulation and then used to estimate energy consumption. However,
a change to the input may produce a new execution path, requiring a new
simulation run to extract the correct instruction sequence. Simulation is
typically several orders of magnitude slower than hardware execution, making
repeated simulations undesirable as a means for tuning or optimizing a program.
ECSA does not depend on repeated simulation. It does not require trace data in
order to provide an energy estimation. This allows for much faster estimation
of a program's energy consumption.

The main difference between energy measurements, trace simulation based energy
predictions and ECSA, is that the first two methods estimate the cost of the
actual executed path. ECSA, however, gives an upper bound based on the cost
model used. Both ECSA and trace estimations rely on the accuracy of the energy
model. Further, they cannot accurately account for energy due to data-sensitive
switching activity. In the rest of this section we will
provide a set of guidelines on how the ECSA results should be interpreted, and
how they can influence energy aware decisions that can be made by software
developers, compiler engineers, development tools and RTOS.

\subsubsection{LLVM IR level ECSA}
\label{LLVMIRuse}

The LLVM optimizer and code emitter are the natural place for compiler optimizations. Our LLVM IR analysis results demonstrate a high accuracy with a deviation in the range of 1\% from the ISA ECSA. Some LLVM IR estimations may not always be as accurate as at ISA level, but they are still of value to developers. Transparency of energy consumption at this level enables programmers to investigate how optimizations affect their program's energy consumption~\cite{TLP:9940890}, or even help introduce new low energy optimizations~\cite{7054192,Pallister:2014}. This is more applicable at the LLVM IR than at the ISA level, because more program information exists at that level, such as types and loop structures. Our mapping techniques and analysis framework at the LLVM IR level are applicable to any compiler that uses the LLVM common optimizer, provided that an energy model for the target architecture is available.

For some programs, indirect jumps that are introduced at the ISA level can make it impossible to extract a CFG. While this prevents ISA level ECSA, the analysis can still be performed for these programs at LLVM IR, allowing programmers to gain energy consumption insight even when ISA level analysis is not feasible.

\subsubsection{ECSA bound use cases}
\label{boundUse}

Given that we are using bound analysis with an energy model characterized with random input data, we must consider the ECSA estimations as loose upper bounds of the WCEC. Although, these bounds are not safe, in most cases they can provide useful information to the programmer, e.g. to determine whether or not an application is likely to exceed an available energy budget.

The modified \texttt{B.Radix4Div} benchmark avoids an early return when the dividend is greater than 255. Omitting this optimization is less efficient, but balances the CFG paths. The effect of this modification can be seen in \Cref{fig:results2}. The ISA level energy consumption lower bounds (the best case retrieved by IPET) are shown. In the optimized version, the energy
consumption across different test cases varies significantly, creating a large range between the upper and lower energy consumption bounds. Conversely, the unoptimized version shows a lower variation, thus narrowing the margin between the upper and lower bounds, but has a higher average energy consumption.

Knowledge of such energy consumption behavior can be of value for applications
like cryptography, where the power profile of systems can be monitored to
reveal sensitive information in side channel attacks~\cite{Kocher:1999}. 
In these situations, ECSA analysis can help code
developers to design code with low energy consumption variation, so that any
potential leak of information that could be obtained from power monitoring can
be obfuscated.

\subsubsection{Parametric resource usage equations}
\label{costFunctions}

Regression analysis was applied to the ISA level static analysis results of the
benchmarks \texttt{MatMult 1,2,4}, \texttt{Mac}, \texttt{Cnt} and
\texttt{Base64}. The resultant upper bound equations are shown in
\Cref{tab:rAnalysis}. The second column shows the retrieved equations which
return the energy consumption predictions in nano-Joules (nJ) as a function over $x$, as defined in the third column. \texttt{Levenshtein} is a multi-parametric energy consumption benchmark. However, the regression analysis was unable to determine a good parametric equation for it.

Parametric resource usage equations can be valuable for a programmer or user to
predict energy consumption with specific parameter values. Moreover, embedding
such equations into an operating system can enable energy aware decisions for
either scheduling tasks, or checking if the remaining energy budget is adequate
to complete a task. If the application permits, the operating system may also
downgrade the quality of service to complete the task within a lower energy
budget.

\begin{table}[h]
\centering
\scriptsize
\begin{tabular}{|l|l|c|l|}
\rowcolor[HTML]{C0C0C0}
\hline
\multicolumn{1}{|c|}{\cellcolor[HTML]{C0C0C0}Benchmark} & \multicolumn{1}{c|}{\cellcolor[HTML]{C0C0C0}Regression Analysis (nJ)} & \multicolumn{1}{c|}{\cellcolor[HTML]{C0C0C0}$x$} \\ \hline
Base64                                                  & $f(x) = 19x + 94.2$                              & string length                                    \\ \hline
Mac                                                     & $f(x) = 15x + 21.1$                              & length of two vectors                      \\ \hline
Cnt                                                     & $f(x) = 19.9x^2 + 5.7x + 34.6$                              & matrix size                                    \\ \hline
MatMul                                                  & $f(x) = 12.2x^3 + 17.5x^2 + 4.7x + 33$                              & size of square matrices                         \\ \hline
MatMul\_2T                                              & $f(x) = 19.3x^3 + 21.4x^2 + 5.9x + 96.8$                              & size of square matrices                         \\ \hline
MatMul\_4T                                              & $f(x) = 22.7x^3 + 25x^2 + 6.5x + 157.7$                              & size of square matrices                                               \\ \hline
\end{tabular}
\caption{\label{tab:rAnalysis}Benchmarks with parametric energy consumption.}
\end{table}
\vspace{-2mm}

\subsubsection{Multi-threaded ECSA}
\label{multi-threaded_usage}

The first class of parallel programs to which ECSA was applied is
replicated non-communicating threads. The user can make energy aware
decisions on the number of threads to use, with respect to time and energy
estimations retrieved by our analysis. For example, take four independent
matrix multiplications on four pairs of equally sized matrices ($28\times28$). Our
analysis will show that a single thread will have an execution time of 4x the
time needed to execute one matrix multiplication. However, two threads will
half the execution time and decrease the energy by 54\%. Four threads which
will half the execution time again, and decrease the energy by 41\% compared to
the two-thread version. Using more threads increases the power dissipation, but
the reduction in execution time saves energy on the platform under
investigation. Although there is a different estimation error between different
numbers of active threads, the error range of 6\% is small enough to allow
comparison between these different versions. The comparison can be also done by
RTOS using the cost functions from \Cref{tab:rAnalysis} to make real
time energy aware scheduling decisions.

The second class of parallel programs that our ECSA was applied to was streaming
pipelines of communicating threads. There is a choice in how to spread the
computation across threads to maximize throughput and therefore minimize
execution time or lower the necessary device operating frequency. Having a
number of available threads, a number of cores and the ability to apply voltage
and frequency scaling, provides a wide range of configuration options in the
design phase, with multiple optimization targets. This can range from
optimizing for quality of service, time and energy, or a combination of all
three. Our ECSA can take advantage of the fact that the energy model used can be
parametric to voltage and frequency, to statically identify the most energy efficient
configuration of the same program, among a number of different options that
deliver the same required performance. The first step of analyzing the
pipelined versions of industrial filter applications has been made in this paper. We
are currently working on extending our ECSA to automatically exploit the
possible different configurations and provide the optimal solution, within the user's constraints.

Finally, the user needs to be aware of the potential effect of input data. When
highly data sensitive applications are analyzed, the user can make some
assumptions, based on the possible input data range, about the accuracy of the
ECSA analysis. As explained in \Cref{dataEffect}, data that is close to random
will lead to a smaller estimation error, when random data was used to build the
energy model. From our findings, this variation can be up to 25\%, but this has only
been shown in short, contrived cases and is unlikely to be large in realistic programs.

\section{Conclusion and Future Work}
\label{sec:conc_future}

This work has given critical review of ECSA existing works that have overlooked the effect of using non data sensitive energy models and SRA bound techniques, on the retrieved energy estimations. In the absence of average case SRA and data sensitive energy models, we establish
this effect in our experimental evaluation of ECSA on a set of mainly industrial benchmarks. We also demonstrate that such an analysis can still have a significant value for software developers, compiler engineers, development tools and RTOS, by establishing a number of ECSA applications in \Cref{ECSAapplications}.

A technique was introduced to allow energy characterization of LLVM IR. It
enables ECSA at this level with a small loss of accuracy, typically 1\%,
compared to ECSA at ISA level. ECSA is applied to a set of multi-threaded
programs for the first time to our knowledge. This is a significant step beyond
existing work that examines single-thread programs, because such an analysis can
provide significant guidance for time-energy design space exploration between
different numbers of threads and cores.

This work has generated new research questions. There is a clear need for non bounding SRA techniques that focus on average cases. Data sensitive energy models and SRA techniques are needed for ECSA to account for data sensitive switching activity in the processor. The majority of existing energy models are usually generated using random data. As we have discussed in \Cref{dataEffect}, alternative data energy models might be better for specific applications.

Future work aims to analyze more complex concurrent programs, such as distinct non-communicating threads rather than replicated threads, pipelines of threads with unbalanced workloads and other concurrency patterns. The ECSA can be combined with some more dynamic techniques such as abstract simulation to account for all the possible threads interleaving. Extending such analysis beyond deeply embedded systems, with more architectural performance enhancing features, might be done by exploiting more techniques from the WCET community, such as abstract interpretation and data cache analysis.

\bibliography{typeinst}
\section*{Acknowledgments}\label{sec:Acknowledgments}
The research leading to these results has received funding from the European
Union 7th Framework Programme (FP7/2007-2013) under grant agreement no 318337,
ENTRA - Whole-Systems Energy Transparency. Special thanks to Intel for providing us with the equipment used for our power monitoring setup.
\end{document}